## Simple approximate analytical solution for non-isothermal single-step transformations: kinetic analysis

J.Farjas* and P.Roura

GRMT, Department of Physics, University of Girona, Campus Montilivi, Edif. PII, E17071 Girona, Catalonia, Spain

### Abstract

In this paper, we develop a method for obtaining the approximate solution for the evolution of single-step transformations under non-isothermal conditions. We have applied it to many reaction models and obtained very simple analytical expressions for the shape of the corresponding transformation rate peaks. These analytical solutions represent a significant simplification of the system's description allowing easy curve fitting to experiment. A remarkable property is that the evolutions of the transformed fraction obtained at different heating rates are identical when time is scaled by a time constant. The accuracy achieved with our method is checked against several reaction models and different temperature dependencies of the transformation rate constant. It is shown that its accuracy is closely related with that of the Kissinger method.

*Corresponding author: jordi.farjas@udg.cat, tel. (34) 972 418490, fax. (34) 972 418098

Nomenclature list

| | |
|---|---|
| α | degree of transformation |
| $\alpha_s$ | transformed fraction that can be chosen at will between 0 and 1, Eq. (9) |
| $\alpha_P$ | transformed fraction at the peak temperature, Eq. (16) |
| β | constant heating rate (K/s) |
| $\Delta t_{FWHM}$ | transformation rate peak width calculated at the full width at half maximum (FWHM). |
| $\Delta t'_{FWHM}$ | transformation rate peak width calculated at the full width at half maximum (FWHM) of the scaled system. |
| $\tau_s$ | scaling factor defined as $\tau_s \equiv 1/k(T_S)$ (s) |
| $\tau_P$ | scaling factor calculated at the peak temperature $\tau_P \equiv 1/k(T_P)$ (s) |
| A | pre-exponential factor of the rate constant in Eq. (2) ($s^{-1}$) |
| $b$ | parameter defined in Eq. (24) ($K^{-1}$) |
| $B$ | parameter defined in Eq. (28) ($K^3$) |
| $B_P$ | constant defined in Eq. (A.7), $B_P \equiv \ln\left(-\frac{AR}{E_A} f'(\alpha_P)\right)$ |
| $B_S$ | constant defined in Eq. (A.6), $B_S \equiv \ln\left(\frac{AR}{E_A\, g(\alpha_S)}\right)$ |
| $E_A$ | activation energy of Eq. (7) (kJ/mol) |
| $f(\alpha)$ | conversion function in Eq. (1) |
| $f'(\alpha)$ | differential of conversion function, $f'(\alpha) \equiv df(\alpha)/d\alpha$ |
| $g(\alpha)$ | function defined by Eq. (3) |
| $G(z)$ | inverse function of $g(\alpha)$ |
| $k(T)$ | rate constant |
| $p(x)$ | temperature integral, Eq. (8) |
| $R$ | universal gas constant (8.314472 J/K·mol) |
| $t$ | time (s) |
| $T$ | absolute temperature (K) |
| $T_0$ | parameter defined in Eq. (24) (K) |
| $T_m$ | parameter defined in Eq. (28) (K) |
| $T_P$ | peak temperature (K) |
| $T_S$ | temperature associated to a given transformed fraction $\alpha_s$ (K) |

$x(T)$ function of temperature in the exponent of Eq. (2)

$x_S'$ temperature derivative of $x(T)$ evaluated at $T_S$.

## Introduction

Solid state transformations are usually described by a single-step kinetic equation:[1]

$$\frac{d\alpha}{dt} = k(T) \cdot f(\alpha) \tag{1}$$

where $\alpha$ is the degree of transformation (0<α<1), $t$ is time, $T$ is the temperature, $k(T)$ is the rate constant and $f(\alpha)$ is the conversion function of a particular reaction model. In Table I, we summarize several published reaction models. Most of them are based on a particular microscopic mechanism[2] (phase boundary reactions, nucleation and growth, diffusion…) except for the nth-order reaction and the Šesták-Berggren equation, which are empirical models.

Solid state transformations are usually thermally activated processes, i.e. the system must overcome an energy barrier. Under this condition, the rate constant can generally be described by

$$k(T) = A\exp\left[-x(T)\right] \tag{2}$$

where $A$ is the pre-exponential factor and $x(T)$ is a function of temperature. In most cases, $A$ is considered to be temperature-independent. This will also be true in the present paper.

The solution of the single-step kinetic equation under isothermal conditions can be obtained by direct integration of Eq. (1):

$$g(\alpha) \equiv \int_0^\alpha \frac{du}{f(u)} = k(T)\left(t - t_0\right), \tag{3}$$

whence

$$\alpha(t) = G\left(k(T)\left(t - t_0\right)\right), \tag{4}$$

where $G$ is a function that is inverse to $g(\alpha)$. From Table I it can be observed that both $g$ and $G$ have an analytical solution for a good number of reaction models, i.e. in most cases one can easily derive an analytical solution for the phase transformation evolution under isothermal conditions.

Eq. (1) is also valid for constant heating conditions.[3] The explicit dependence on time of Eq. (1) can be easily eliminated:

$$\frac{d\alpha}{dT} = \frac{1}{\beta} k(T) \cdot f(\alpha) \tag{5}$$

where $\beta = dT/dt$ is the constant heating rate. Integration of Eq. (5) gives the solution of the single-step kinetic equation under continuous heating conditions:

$$g(\alpha) = \frac{1}{\beta}\int_{T_0}^{T} k(T)dT \tag{6}$$

The right hand term in Eq. (6) is the temperature integral[4] and, unlike the isothermal solution, it does not have a simple analytical solution.

Most solid state transformations are properly described by an Arrhenius dependence of the rate constant:[5-7]

$$k(T) = A\exp\left(-\frac{E_A}{RT}\right) \tag{7}$$

where $E_A$ is the activation energy and $R$ is the universal gas constant ($E_A$ is temperature-independent). With this particular temperature dependence and supposing that the transformation rate at $T_0$ is negligible, Eq. (6) can be rewritten as

$$g(\alpha) = \frac{E_A A}{\beta R}\, p\left(\frac{E_A}{RT}\right) \tag{8}$$

where $p(x) \equiv \int_{\infty}^{x} \frac{\exp(-u)}{u^2} du$.[4,8] Although this is an exact analytical solution of the temperature integral, it requires numerical integration. Consequently, the properties of $\alpha(T)$ for particular models can only be deduced with great effort. To obtain simpler analytical solutions, several authors have developed approximate solutions of the previous temperature integral .[4, 9-16]

In this paper, we develop an approximate analytical solution of the transformed fraction evolution by assuming that the transformation takes place in a narrow temperature range (Section 2). This approach has already been successfully applied in the case of solid phase transformations ruled by the Kolmogorov-Johnson-Mehl-Avrami (KJMA) model.[17] In Section 3, we verify the validity of our approach for two different reaction models when the rate constant follows an Arrhenius dependence. In Section 4, we test the validity of our approach for different temperature dependences of the rate constant and, finally, in Section 5, we apply our approach to the experimental crystallization curves of amorphous silicon.

## Approximate solution under continuous heating conditions

If the transformation takes place in a narrow temperature range, the function $\ln[g(\alpha)]$ can be substituted by its first order series expansion in temperature:

$$\ln[g(\alpha(T))] \approx \ln[g(\alpha_S)] + \frac{k(T_S)}{\beta g(\alpha_S)}(T - T_S), \tag{9}$$

where $\alpha_s$ is a reference that can be chosen at will between 0 and 1. In addition, if we choose the time origin such that $\alpha(t=0) = \alpha_s$, then $T = T_s + \beta t$. Thus, Eq. (9) becomes

$$\ln[g(\alpha(t))] \approx \ln[g(\alpha_S)] + \frac{1}{g(\alpha_S)}\frac{t}{\tau_S} \quad , \tag{10}$$

where $\tau_s \equiv 1/k(T_S)$. Eq. (10) can be transformed to

$$g(\alpha) = g(\alpha_S) \cdot \exp\left(\frac{1}{g(\alpha_S)}\frac{t}{\tau_S}\right). \tag{11}$$

Now, the analytical solution for a particular kinetic model is obtained by inverting the $g(\alpha)$ function:

$$\alpha(t) = G(z), \text{ with } z \equiv g(\alpha_S)\exp\left(\frac{1}{g(\alpha_S)}\frac{t}{\tau_S}\right) \tag{12}$$

The inverse $G$ functions for a number of models are detailed in Table I. Finally, the transformation rate can be obtained from the time derivative of Eq. (11):

$$\frac{d\alpha}{dt} = f(\alpha)\frac{1}{\tau_S}\exp\left(\frac{1}{g(\alpha_S)}\frac{t}{\tau_S}\right) = f(\alpha)\frac{1}{g(\alpha_S)\tau_S}\exp(z) \tag{13}$$

It is worth noting that this solution is scalable with time by a scaling factor $\tau_s$ and that the scaled solution does not depend on $\beta$ or the rate constant.

From an experimental point of view, it is particularly interesting to choose $\alpha_s$ as the value at which the maximum transformation rate is achieved. Henceforth we will note this particular value as $\alpha_P$. The corresponding temperature is the so-called peak temperature, $T_P$. The value of $\alpha_P$ can be deduced from the first derivative of Eq. (13),

$$\frac{d^2\alpha}{dt^2} = f'(\alpha)\frac{d\alpha}{dt}\frac{1}{\tau_S}\exp\left(\frac{1}{g(\alpha_S)}\frac{t}{\tau_S}\right) + f(\alpha)\frac{1}{\tau_S^2 g(\alpha_S)}\exp\left(\frac{1}{g(\alpha_S)}\frac{t}{\tau_S}\right), \tag{14}$$

and imposing that $\left.\frac{d^2\alpha}{dt^2}\right|_{T_P} = 0$:

$$f'(\alpha_P)\exp\left(\frac{1}{g(\alpha_S)}\frac{t_P}{\tau_S}\right) = -\frac{1}{g(\alpha_S)} \tag{15}$$

Since $\alpha_S$ can be chosen at will, for $\alpha_S = \alpha_P$, $t_P$=0, Eq. (15) becomes

$$-f'(\alpha_P)g(\alpha_P) = 1. \qquad (16)$$

Actually, the previous relation is a well-known result[33,34] that has been deduced for an Arrhenius temperature dependence of the rate constant. In fact, if we assume an Arrhenius dependence, the first derivative of Eq. (1) becomes,

$$\frac{d^2\alpha}{dt^2} = k(T)\frac{\beta E_A}{RT^2} f(\alpha) + k(T) f'(\alpha)\frac{d\alpha}{dt} = k(T) f(\alpha)\left[k(T) f'(\alpha) + \frac{\beta E_A}{RT^2}\right] \qquad (17)$$

then imposing the condition of the peak temperature,

$$f'(\alpha_P)\frac{ART_P^2}{\beta E_A}\exp\left(-\frac{E_A}{RT_P}\right) = -1 \qquad (18)$$

and inserting the approximation of Murray and White[9] $\left(p(x) \approx \frac{\exp(-x)}{x^2}\right)$ in Eq. (8),

$$g(\alpha) = \frac{E_A A}{\beta R} p\left(\frac{E_A}{RT_P}\right) \approx \frac{ART_P^2}{\beta E_A}\exp\left(-\frac{E_A}{RT_P}\right) \qquad (19)$$

Finally, if we substitute Eq. (19) into Eq. (18) one obtains Eq. (16).

From Eq. (16) it can be concluded that within our approximation, $\alpha_P$ does not depend on $\beta$ or the rate constant. Therefore, the peak temperature is related to a fixed transformed fraction. This conclusion is as expected since the solution (Eq. (12)) is scalable with time. For several kinetic models Gao et al.[33] showed that when $E_A / RT_P > 20$ the values of $\alpha_P$ indeed fall into a very narrow range. Experience shows that this condition is fulfilled for most solid phase transformations. Note that the value of $\alpha_P$ given by Eq. (16) corresponds to the value of $\alpha_P$ in the limit of $E_A / RT_P \to \infty$.[33,34] In Table I we have included this value of $\alpha_P$ as well as that of $g(\alpha_P)$ for several kinetic models. Observe that for most kinetic models $g(\alpha_P) = 1$. Thus, selecting $\alpha_P$ as a reference generally makes Eq. (12) simpler.

**Validity check for the approximate solution: Arrhenius dependence**

In this section, we will assume that the rate constant follows an Arrhenius dependence (Eq. (7)), which is, by far, the most commonly used equation to describe solid phase transformations. Two different particular cases, which are described by different kinetic models, will be analyzed. In addition to these two models, the validity of our approximation has already been verified for a particular case of the KJMA(n) model.[17]

*Decomposition of calcium carbonate*

First, we will focus our attention on the decomposition of calcium carbonate under nitrogen. We have chosen this reaction because it has been studied in depth and it follows a single-step kinetics fairly well.[35-38] The decomposition of calcium carbonate can be described by two different models: Rn with $n$=1.215 and Fn with $n$=0.177. From Table I, one can easily verify that both models are mathematically equivalent. The kinetic parameters obtained under continuous heating are ln($A$)=15.86 ($A$ in $s^{-1}$) and $E_A$=194.26 kJ $mol^{-1}$.[38] In Fig. 1 we show the exact transformation and transformation rate curves ($\alpha(T)$ and $d\alpha/dt$ respectively) obtained from the numerical integration of Eq. (1) for this particular case at different heating rates. The approximate solution is obtained after substituting the corresponding $G$ function (Table I) into Eq. (12). With the specific choice of $\alpha_S = \alpha_P$, we get ($\alpha_S = 1 - n^{1/(n-1)}$ , $g(\alpha_S) = 1$):

$$\alpha(t) = 1 - \left[1 + (n-1)\, t / \tau_P\right]^{1/(1-n)} \ . \tag{20}$$

First of all, we will check the validity of the scaling law. Since the transformed fraction obtained at different heating rates are identical when time is scaled by $\tau_S$, the width of the peak must be proportional to $\tau_S$. To quantify the peak width we have calculated the full width at half maximum (FWHM), $\Delta t_{FWHM}$;

$$\Delta t_{FWHM} = \tau_S \Delta t'_{FWHM} , \tag{21}$$

where $\Delta t'_{FWHM}$ is the FWHM of the scaled system, i.e. a constant that only depends on the particular kinetic model and that is independent of the rate constant. $\Delta t'_{FWHM}$ can be obtained from Eqs. (20) and (13) by substituting $t$ with the scaled time $t' \equiv t/\tau_S$ ; the resulting value is 1.17865 for $n$=0.177.

Taking into account the definition of $\tau_P$ one obtains

$$\ln\left(\Delta t_{FWHM}\right) = \frac{E_A}{R\,T_P} + \ln\left(\frac{\Delta t'_{FWHM}}{A}\right). \tag{22}$$

Thus the plot of $\ln\left(\Delta t_{FWHM}\right)$ versus $1/T_P$ must be a straight line with a slope equal to $E_A/R$. In Fig. 2 we have plotted $\ln\left(\Delta t_{FWHM}\right)$ for the exact transformation rate peak versus $1000/T_P$ (symbols). One can observe that the exact values of $\Delta t_{FWHM}$ are perfectly aligned. The slope of the fitting line is 23340 K and the y-intercept value is -15.65. These results are in very good agreement with the values predicted by our approximate solution, which are 23360 K and -15.69 (Table II). In Fig. 2 we have also included the Kissinger plot of the exact peak temperature (Eq. (A.7)) which fits a straight line with a slope of -23310 K and y-intercept 5.68. These values also agree with the theoretical prediction (Table III).

Finally, in Fig. 3 we compare the transformation curves $\alpha$ and $d\alpha / dt$ given by our approximate solution, Eq. (20), to those obtained by exact integration of Eq. (1) for two extreme heating rates of $\beta$=0.005 and 100 K/min, which can be considered to be the lower and upper limits for most experiments. The curves are plotted versus the scaled time, $t / \tau_P$. The coincidence between the two heating rates and the approximate solution is excellent. The discrepancies in the transformed fraction between the analytical solution and the exact result for the two heating rates are lower than 2 $10^{-2}$ despite the large shift in peak temperatures from 545.4 to 932.2 ºC and the very different time-scaling factors of 3.22 $10^{5}$ and 34.0 s for $\beta =$ 0.005 and 100 K/min respectively. It is worth noting that, although the two time scales differ by four orders of magnitude, the scaling law is valid. In fact, the usefulness of our approximation is based on the Arrhenius dependence of the rate constant. This strong dependence on temperature limits the transformation to a narrow temperature range even when the heating rate is as low as 0.005 K/min.

*Reduction of nickel oxide*

As a second test we have selected a transformation which fits Šesták-Berggren kinetics. To work with realistic parameters we have taken the parameters obtained from the kinetic analysis of nickel oxide reduction in a hydrogen atmosphere:[39] $m$=0.63, $n$=1.39, $p$=0, ln($A$)=15.4 (A in $s^{-1}$) and $E_A$=96.4 kJ $mol^{-1}$. Although not shown, the plot of $\Delta t_{FWHM}$ versus 1000/$T_P$ and the Kissinger plot for $\beta$ ranging from 0.005 to 100 K/min exhibit a very linear trend and the calculated slope and y-intercept agree with the theoretical values (see Tables II and III).

In this case, no side of Eq. (6) has an analytical solution. However, our solution still represents a significant simplification since, according to Eq. (11), calculating the evolution of the transformed fraction can be reduced to calculating $g(\alpha)$ from its definition, Eq. (3), and then evaluating the related time:

$$t = \tau_S \ln\left[\frac{g(\alpha)}{g(\alpha_S)}\right] \tag{23}$$

Note also that, thanks to the time scaling property, the calculation has to be done only once. The particular evolution for a given heating rate is obtained by multiplying the scaled time by the corresponding $\tau_S$. In this case, we do not have any analytical expression for $\alpha_P$. Therefore, we have selected a different reference, $\alpha_S = m/(m+n)$, which corresponds to the maximum of $f(\alpha)$.[34] In Fig. 4 we have plotted the exact values of $\alpha(t)$ and $d\alpha/dt$ (symbols) for two extreme heating rates of $\beta$=0.005 and 100 K/min. Here, again, the coincidence between the two heating rates and the approximate solution (solid lines) is excellent despite the large shift in peak temperatures from 436 to 662 K and the very different time-scaling factors of 9.76 $10^4$ and 10.8 s for $\beta =$ 0.005 and 100 K/min respectively.

## Check of the validity of the approximate solution: Non-Arrhenius behavior

In this section we will apply our approximation to solid-state reactions in which the temperature dependence of the rate constant is not an Arrhenius one.

### *Crystallization of $Pd_{80}Si_{20}$*

The rate constant for the crystallization of glasses generally exhibits an Arrhenius or Vogel-Fulcher temperature dependence.[40] The Vogel-Fulcher expression of the rate constant is:

$$k(T) = A\exp\left(-\frac{1}{b(T-T_0)}\right) \tag{24}$$

where $b$ and $T_0$ are empirical parameters. As a reference, we have taken the kinetic parameters obtained from the crystallization of the metallic glass $Pd_{80}Si_{20}$:[41] $A$=6 $10^6$ $s^{-1}$, $b$=2 $10^{-4}$ $K^{-1}$ and $T_0$=428 K. The kinetic model is the KJMA(n) with $n$=3.

According to Table I and Eq. (12) the approximate solution is

$$\alpha(t) = 1 - \exp\left[-\left(\exp[t/\tau_P]\right)^n\right] \tag{25}$$

Bearing in mind the definition of $\tau_S$ and substituting Eq. (24) into Eq (21) one obtains for a Vogel-Fulcher temperature dependence:

$$\ln\left(\Delta t_{FWHM}\right) = \frac{1}{b\left(T_P - T_0\right)} + \ln\left(\frac{\Delta t'_{FWHM}}{A}\right) \tag{26}$$

Where $\Delta t'_{FWHM} = 2.44639/n$ for the KJMA(n) model.[17] In addition, an equivalent Kissinger plot can be obtained after combining Eqs. (24), (16) and (A.3) (see Appendix A):

$$\ln\left(\frac{\beta}{(T_P - T_0)^2}\right) = -\frac{1}{b(T_P - T_0)} + \ln\left(-Ab\, f'(\alpha_P)\right) \tag{27}$$

Similarly to the previous cases, for $\beta$ ranging from 0.005 to 100 K/min, we have plotted $\ln\left(\Delta t_{FWHM}\right)$ versus $1/(T_P\text{-}T_0)$ and a plot equivalent to the Kissinger plot. As expected, both representations give a straight line. The results of the linear fitting are summarized in Tables II and III. One can verify that the fitted parameters agree with the predicted values, Eqs. (25) and (26). The agreement is especially remarkable for $\Delta t_{FWHM}$.

In Fig. 5 the approximate solution is compared to the exact solution, Eq. (1), calculated numerically at the two extreme heating rates. In this case, the agreement between the approximate solution and the exact one is the best reported in this paper (the discrepancies in $\alpha$ are smaller than $10^{-3}$). This result agrees with Table II, in which the best fitting for $\Delta t_{FWHM}$ is also obtained for this case. Note that again there is a large difference in the time scale: 8.59 s at 100 K/min and 77544 s at 0.005 K/min.

*Crystallization of polyethylene glycol*

We will end this section with an analysis of crystallization from the melt during cooling. For low undercooling the rate constant can be described approximately by:[42]

$$k(T) = A\exp\left(-\frac{B}{T\left(\Delta T\right)^2}\right), \quad \Delta T = T_m - T \tag{28}$$

where $B$ is constant, $A$ can be considered constant in a relatively narrow temperature interval and $T_m$ is the melting temperature. We have selected the kinetic parameters obtained experimentally from the crystallization of polyethylene glycol (PEG):[43] $B$=20256 $K^3$, $T_m$=45.4ºC and ln($A$)=-1.79 ($A$ in $s^{-1}$). The experimental data approximates a Šesták-Berggren model with exponents $n$=1.4, $m$=0.49 and $p$=0.

In Fig. 6 we have plotted $\alpha(t)$ and $d\alpha/dt$ calculated from the exact solution, Eq. (1), and from the approximate solution, Eq. (23), at the two extreme heating rates.

For the approximate solution, we have selected as a reference $\alpha_S = m/(m+n)$. In contrast with the previous cases, there is no agreement between the exact solution and the approximate one, the discrepancies are especially large at -100 K/min. Thus, in this case our approximate solution is quite inaccurate.

From the analysis of the limits of the approximate solution developed in Appendix A, it is clear that the accuracy of our approximation is directly related to the validity of the Kissinger method (Eq. (A3)). In Table IV, it is shown that the condition deduced in Appendix A $g(\alpha_S) \approx -k(T_S)/(\beta x_S')$ is fulfilled by the first three cases analyzed previously but does not work for the present case. The discrepancy is especially large when the heating rate is -100 K/min. Finally, in Fig 7 we have plotted $\ln(\Delta t_{FWHM})$ versus 1000/$T_P$ and the equivalent Kissinger plot for $\beta$ ranging from -0.005 to -100 K/min. The equivalent Kissinger plot can be obtained by combining Eqs. (28), (16) and (A.3):

$$\ln\left(\frac{\beta(T_m - 3T_P)}{T_P^2(T_m - T_P)^3}\right) = -\frac{B}{T(T_m - T_P)^2} + \ln\left(-\frac{A\, f'(\alpha_P)}{B}\right) \quad (29)$$

The Kissinger plot shows a clear deviation from linearity when $|\beta|$ is high. For low $|\beta|$, where our approximate solution approaches the exact behavior (Fig. 6), the linear dependence in the Kissinger plot is recovered (Fig. 7). Furthermore, this linear region has been fitted to a straight line whose parameters deviate less than 10% from the predicted values (Table III). In Appendix A, we show that this correspondence between the accuracy of our solution at the accuracy of the Kissinger plot to deliver the correct kinetic parameters is not fortuitous. From a practical point of view, we can affirm that our approximation fails whenever the Kissinger plot shows significant deviations from linearity. In this context, it is surprising to observe that the plot of $\ln(\Delta t_{FWHM})$ follows a straight line even for large values of $|\beta|$. From Fig. 7 one can ascertain that despite the different shape of the approximate and exact curves obtained at -100 K/min, their widths are similar.

**Analysis of experimental crystallization curves of amorphous silicon**

It is widely accepted that model-free isoconversional methods are the most reliable for obtaining the activation energies of thermally activated reactions.[7,15,37,44] The Kissinger method[45] can be considered as isoconversional for single-step transformations. Thanks to its simplicity, the Kissinger method is probably the most widespread method used for analyzing experimental data. Its main drawback is that, even in the case of complex mechanisms or for transformations involving more than one step, one can obtain a good fitting from which an average activation energy can be deduced. Consequently, from the Kissinger plot, one cannot conclude whether the transformation follows a single-step or a more complex process.[46] This problem is related to the fact that the Kissinger method is based exclusively on the peak temperature, a parameter which is quite insensitive to alterations in the peak shape. Analyzing the peak temperature and $\ln(\Delta t_{FWHM})$ together, which is possible with our approximation, provides a better test of the validity of the Kissinger method. Since $\Delta t_{FWHM}$ is very sensitive to the peak shape, the occurrence of complex processes will result in inconsistencies between the two analyses. These points will be made clearer by analyzing experimental data of amorphous silicon crystallization.

In Fig. 8 we have represented the Kissinger plot and $\ln(\Delta t_{FWHM})$ versus $1/T_P$ obtained from the crystallization of thick films of amorphous silicon.[47] Data have been obtained from differential scanning calorimetric measurements performed at different heating rates. The activation energy obtained from the plot of $\ln(\Delta t_{FWHM})$, 343 kJ/mol, agrees with that obtained from the Kissinger plot, 346 kJ/mol. Note that the agreement between both analyses considerably increases the confidence in the results obtained from the Kissinger analysis; however, it does not represent a definitive proof. Thus, correlation between fitted and physical parameters requires ancillary data. Electron microscopy measurements taken on partially crystallized films[48,49] have determined that amorphous silicon crystallization is ruled by nucleation and growth and follows KJMA kinetics. The activation energies for homogeneous nucleation and growth are $E_N$=511 and $E_G$=299 kJ/mol respectively. In the case of thick silicon films, three dimensional growth is expected. It has been recently stated[24] that under these circumstances the crystallization rate can be described with Eq. (1) where $E_A = (E_N + 3E_G)/4 = 352$ kJ/mol and $f(\alpha)$ corresponds to the KJMA model with $n$=4. Note that the agreement between calorimetry and microscopy confirms that the observed crystallization kinetics corresponds to a process ruled by homogeneous nucleation and three dimensional growth.

A kinetic analysis method should also provide a kinetic model which allows the transformation to be correctly described. Without auxiliary data, one cannot expect to be able to determine the specific mechanism of the transformation solely from the kinetic analysis. However, one should expect coherence between the rate constants and the kinetic model. According to our approximation, not only $\Delta t_{FWHM}$ but the whole shape of the transformation rate should scale with time. Thus, once the parameters of the rate constant have been determined (Fig. 8), a plot of the transformation rate versus the scaled time should give a universal curve, i.e. a curve independent of the heating rate. If fulfilled, the latter condition, which involves all the experimental data, clearly reinforces the fact that the measured kinetics corresponds mainly to a single-step transformation and ensures the coherence between the rate constants and the kinetic model. Furthermore, from the universal curve one can obtain the kinetic model using the standard fitting techniques.

Alternatively, one can test a particular kinetic model by calculating our approximate solution, Eq. (12), and comparing it to the experimental curves obtained at different heating rates by substituting the corresponding value of $\tau_P$ obtained from the analysis of the rate constant. We have selected this procedure in the case of the crystallization of amorphous silicon. In particular, we tested the KJMA model, Eq. (27), with *n*=4. The result is plotted in Fig. 9. The value of $\tau_P$ was obtained directly from the plot of $\ln(\Delta t_{FWHM})$ taking into account that $\Delta t'_{FWHM} = 2.44639/n$ for the KJMA(n) model.[17] Comparing experimental data and simulated values in Fig. 9 clearly confirms that the approximate solution gives a good description, and that the kinetics indeed corresponds to homogeneous nucleation and growth. Note that the thermogram measured at 14 K/min exhibits a significant discrepancy from the predicted shape. This deviation corresponds to a quicker heterogeneous nucleation process that takes place on the surface of the films. This secondary process is best resolved as a small sharp peak at lower heating rates (notably below 1.0 K/min). However, we consider that the failure to correctly fit the 0.25 K/min curve is due to the difficulty of resolving the weak signal from an unstable baseline.

## Conclusions

We have developed a simple method that allows obtaining simple approximate analytical solutions for the kinetics of single-step transformations under non-isothermal conditions, Eq. (12). These solutions are scalable in time and the scaled solutions do not depend on the heating rate or the rate constant. The correctness of our method is directly related to the applicability of the Kissinger analysis, i.e. it can be applied to all the single-step transformations which obey the Kissinger relation.

The solutions detailed for a number of reactions models, Table I, can be directly applied to the analysis of experimental curves. Their simplicity allows curve fitting without need of solving any differential equation and, in most cases, without numerical integration. In addition, the existence of the existence of a universal scaled solution constitutes a very restrictive test of the assumption of single-step transformation that can be used to reveal complexities in the reaction kinetics

## Acknowledgments

This work was supported by the Spanish *Programa Nacional de Materiales* under contract number MAT2006-11144.

## Appendix A. Analysis of the accuracy of the approximate solution

Our approximate solution is accurate provided that the second term in the Taylor development of Eq. (9) is negligible when compared to the first term:

$$\frac{k(T_S)}{\beta\, g(\alpha_S)}\Delta T >> \frac{1}{2}\frac{k(T_S)}{\beta\, g(\alpha_S)}\left|\frac{k(T_S)}{\beta\, g(\alpha_S)} + x_S'\right|\Delta T^2 \qquad \text{(A.1)}$$

$\Delta T$ is the temperature range in which the transformation takes place and $x_S' \equiv dx/dT\big|_{T=T_S}$. Rearranging Eq. (A.1) and assuming that $\Delta T \approx 2\beta\,\Delta t_{FWHM} = 2\beta\,\Delta t'_{FWHM}/k(T_S)$, one obtains

$$\frac{\Delta t'_{FWHM}}{g(\alpha_S)}\left|1 + \frac{g(\alpha_S)\,\beta\, x_S'}{k(T_S)}\right| << 1 \qquad \text{(A.2)}$$

Note that the $\Delta t'_{FWHM}/g(\alpha_S)$ term only depends on the kinetic model, i.e. it is independent of the rate constant and the heating rate. In all the cases analyzed, this parameter is of the order of 1 (see Table IV). Thus, our approximate solution is accurate if the following condition is fulfilled:

$$g(\alpha_S) \approx -\frac{k(T_S)}{\beta\, x_S'}. \qquad \text{(A.3)}$$

As shown below, this condition is the basis of the well-known Kissinger-Akahira-Sunose (KAS) or generalized Kissinger method.[50-52]

If the rate constant follows an Arrhenius behavior,

$$x_S' = -\frac{E_A}{RT_S^2} \quad . \qquad \text{(A.4)}$$

(A.3) becomes

$$g(\alpha_S) = \frac{T_S^2}{\beta}\frac{R}{E_A} A \exp\left(-\frac{E_A}{RT_S}\right) \qquad \text{(A.5)}$$

Rearranging Eq. (A.5) and taking logarithms one obtains the KAS relation:

$$\ln\left(\frac{\beta}{T_S^2}\right) = -\frac{E_A}{RT_S} + B_S \qquad \text{(A.6)}$$

where $B_S \equiv \ln\left(\frac{AR}{E_A\, g(\alpha_S)}\right)$ is constant. Thus, for a fixed value of the transformation fraction $\alpha_S$ the plot of $\ln(\beta/T_S^2)$ versus $1/T_S$ results in a straight line from which the activation energy can be obtained.

Furthermore, if we substitute Eq. (16) into Eq. (A.6) we obtain

$$\ln\left(\frac{\beta}{T_P^2}\right) = -\frac{E_A}{RT_P} + B_P \tag{A.7}$$

where $B_P \equiv \ln\left(\frac{AR}{E_A} f'(\alpha_P)\right)$. Eq. (A.7) is the well-known linear relation of the Kissinger method.[45,53] Note that the validity of the Kissinger method relies on the fact that $B_P$ does not depend on $\beta$. Within our approximation $f'(\alpha_P)$ does not depend on $\beta$ or the rate constant, and the Kissinger method can be considered as isoconversional.[15]

**References**

1. Brown ME, Dollimore D, Galwey AK. Theory of solid state reaction kinetics. In: Bamford CH, Tipper CFH, Editors, Chemical Kinetics, Vol. 22. Amsterdam: Elsevier, 1980:41-113.

2. Šesták J, Berggren G. Study of the kinetics of the mechanism of solid-state reactions at increasing temperatures. *Thermochim. Acta.* 1971;3:1-12.

3. Tang TB, Chaudhri MM. Analysis of dynamic kinetic data from solid-state reactions. *J. Thermal. Anal.* 1980;18:247-261.

4. Coats AW, Redfern JP. Kinetic parameters from thermogravimetric data. *Nature*. 1964;201:68-69.

5. Raghavan V, Cohen M. Solid-state transformations. In: Hannay NB. Treatise of solid state chemistry, Vol. 5, Changes of State. New York: Plenum Press, 1975:67-127.

6. Le Claire AD. Diffusion. In: Hannay NB. Treatise of solid state chemistry, Vol. 4, Reactivity of Solids. New York: Plenum Press, 1975:1-59.

7. Vyazovkin S, Wight CA. Isothermal and nonisothermal reaction kinetics in solids: In search of ways toward consensus. *J. Phys. Chem. A*. 1997; 101: 8279-8284.

8. Šesták J. Thermophysical properties of solids, their measurements and theoretical analysis. Amsterdam: Elsevier 1984.

9. Murray P, White J. Kinetics of the thermal dehydration of clays. IV. Interpretation of the differential thermal analysis of the clay minerals. *Trans. Brit. Ceram. Soc*. 1955;54:204-238.

10. Doyle CD. Series approximations to equation of thermogravimetric data. *Nature*. 1965;207:290-291.

11. Senum GI, Yang RT. Rational approximations of integral of Arrhenius function. *J. Thermal. Anal.* 1977;11:445-449.

12. Heal GR. Evaluation of the integral of the Arrhenius function by a series of Chebyshev polynomials - use in the analysis of non-isothermal kinetics. *Thermochim. Acta*. 1999;340/341: 69-76.

13 Woldt E. The relationship between isothermal and nonisothermal description of Johnson-Mehl-Avrami-Kolmogorov kinetics. *J. Phys. Chem. Solids*. 1992;53:521-527.

14. Starink MJ. A new method for the derivation of activation energies from experiments performed at constant heating rate. *Thermochim. Acta*. 1996;288:97-104.

15. Starink MJ. The determination of activation energy from linear heating rate experiments: a comparison of the accuracy of isoconversion methods. *Thermochim. Acta*. 2003;404:163-176.
16. Orfao JJM. Review and evaluation of the approximations to the temperature integral. *AICHE J*. 2007;53:2905-2915.
17. Farjas J, Roura P. Solid phase crystallization under continuous heating: kinetic and microstructure scaling laws. *J. Mat. Res*. 2008;23:418-424.
18. Jacobs PWM, Tompkins FC. Classification and Theory of Solid Reactions. In: Garner WE, Editor, Chemistry of the solid state. London: Butterworths Scientific Publications, 1955:184-212.
19. Avrami M. Kinetics of phase change I - General theory. *J. Chem. Phys*. 1939;7:1103-1112.
20. Avrami M. Kinetics of Phase Change. II Transformation-Time Relations for Random Distribution of Nuclei. *J. Chem. Phys*. 1940;8:212-224.
21. Avrami M. Granulation, Phase Change, and Microstructure - Kinetics of Phase Change. III. *J. Chem. Phys*. 1940;9:177-184.
22. Johnson WA, Mehl RF. Reaction kinetics in processes of nucleation and growth. *Trans. Amer. Inst. Min. Met. Eng*. 1939;135: 416-442.
23. Kolmogorov AN. On the Statistical Theory of Metal Crystallisation. *Izv. Akad. Nauk. SSSR, Ser. Fiz*. 1937;1:355-359.
24. Farjas J, Roura P. Modification of the Kolmogorov-Johnson-Mehl-Avrami rate equation for non-isothermal experiments and its analytical solution. *Acta. Mater*. 2006;54: 5573-5579.
25. Henderson DW. Experimental-analysis of nonisothermal transformations involving nucleation and growth. *J. Thermal Anal*. 1979;15:325-331.
26. Christian JW. The theory of transformation in metals and alloys, part I (3rd edition). Oxford: Elsevier, 2002.
27. Mampel KL. Time-conversion formulae for heterogenous reactions in phase limits of solid bodies - 2 The time-conversion formulae for a powder from globular particles. *Physik. Chem. A*. 1940;187: 235-249.
28. Prout EG, Tompkins FC. The thermal decomposition of potassium permanganate. *Trans. Faraday Soc*. 1944; 40:488-497.
29. Frank-Kamenetskii DA. Diffusion and heat transfer in chemical kinetics. New York: Plenum Press, 1969.

30. Valensi G. Kinetics of oxidation of metallic wires. *Compt. Rend.* 1935;201:602-604.
31. Jander W. Reactions in solid states at room temperature I Announcement the rate of reaction in endothermic conversions. *Z. Anorg. Allgem. Chem.* 1927;163:1-30.
32. Ginstling AM, Brounshtein BI. O diffuzionnoi kinetike reaktsii v sfericheskikh chastitsakh. *Zh. Prikl. Khim.* 1950;23: 1249-1259. (English translation p. 1327)
33. Gao X, Chen D, Dollimore D. The correlation between the value of alpha at the maximum reaction-rate and the reaction-mechanisms - a theoretical-study. *Thermochim. Acta*. 1993; 223:75-82.
34. Málek J. The kinetic-analysis of nonisothermal data. *Thermochim. Acta*. 1992;200:257-269.
35. Gallagher PK, Johnson DW. The effects of sample size and heating rate on the kinetics of the thermal decomposition of $CaCO_3$. *Thermochim. Acta*. 1973;6:67-83.
36. Salvador AR, Garcia Calvo E, Aparicio CB. Effects of sample weight, particle size, purge gas and crystalline structure on the observed kinetic parameters of calcium carbonate decomposition. *Thermochim. Acta*. 1989;143:339-345.
37 Brown ME, Maciejewski M, Vyazovkin S, Nomen R, Sempere J, Burnham A, Opfermann J, Strey R, Anderson HL, Kemmler A, Keuleers R, Janssens J, Desseyn HO, Li CR, Tang TB, Roduit B, Malek J, Mitsuhashi T. Computational aspects of kinetic analysis Part A: The ICTAC kinetics project-data, methods and results. *Thermochim. Acta*. 2000;355:125-143.
38. Roduit B. Computational aspects of kinetic analysis. Part E: The ICTAC Kinetics Project - numerical techniques and kinetics of solid state processes. *Thermochim. Acta*. 2000;355:171-180.
39. Janković B, Adnađević B, Mentus S. The kinetic analysis of non-isothermal nickel oxide reduction in hydrogen atmosphere using the invariant kinetic parameters method. *Thermochim. Acta*. 2007;456:48-55.
40. Henderson DW. Thermal-analysis of nonisothermal crystallization kinetics in glass forming liquids. *J. Non Crys Solids*. 1979;30:301-315.
41. Telleria I, Barandiaran JM. Kinetics of the primary, eutectic and polymorphic crystallization of metallic glasses studied by continuous scan methods. *Thermochim. Acta*. 1996; 280/281:279-287.
42. Clavaguera N, Saurina J, Lheritier J, Masse J, Chauvet A, ClavagueraMora MT. Eutectic mixtures for pharmaceutical applications: A thermodynamic and kinetic study. *Thermochim. Acta*. 1997; 290:173-180.

43. Berlanga R, Farjas J, Saurina J, Suñol JJ. A modified method for T-CR-T diagram construction - Application to polyethylene glycol. *J. Thermal. Anal.* 1998;52:765-772.
44. Burnham AK. Computational aspects of kinetic analysis. Part D: The ICTAC kinetics project - multi-thermal-history model-fitting methods and their relation to isoconversional methods. *Thermochim. Acta*. 2000; 355:165-170.
45. Kissinger HE. Reaction kinetics in differential thermal analysis. *Anal. Chem*. 1957;29:1702-1706.
46. Sbirrazzuoli N, Girault Y, Elegant L. Simulations for evaluation of kinetic methods in differential scanning calorimetry 3. Peak maximum evolution methods and isoconversional methods. *Thermochim. Acta*. 1997; 293:25-37.
47. Farjas J, Rath C, Roura P, Cabarrocas PR. Crystallization kinetics of hydrogenated amorphous silicon thick films grown by plasma-enhanced chemical vapour deposition. *Appl. Surf. Sci*. 2004;238:165-168.
48. Iverson RB, Reif R. Recrystallization of amorphized polycrystalline silicon films on $SiO_2$ - temperature-dependence of the crystallization parameters. *J. Appl. Phys*. 1987;62:1675-1681.
49. Spinella C, Lombardo S, Priolo F. Crystal grain nucleation in amorphous silicon. *J. Appl. Phys*. 1998;84:5383-5414.
50. Sunose T, Akahira T. Method of determining activation deterioration constant of electrical insulating materials. *Research Report, Chiba Inst. Tecnol. (Sci. Tecnol.)* 1971; 16:22-23.
51. Mittemeijer EJ. Analysis of the kinetics of phase-transformations. *J. Mater. Sci.* 1992;27:3977-3987.
52. Starink MJ. On the applicability of isoconversion methods for obtaining the activation energy of reactions within a temperature-dependent equilibrium state. *J. Mater. Sci*. 1997;32: 6505-6512.
53. Yinnon H, Uhlmann DR. Applications of thermoanalytical techniques to the study of crystallization kinetics in glass-forming liquids .1. Theory. *J Non-Cryst Solids*. 1983;54:253-275.

**Table I**. Set of kinetic models, $f(\alpha)$, integral form of kinetic models, $g(\alpha)$, function inverse to $g(\alpha)$, *G(z)*, approximate value of $\alpha$ at the peak temperature, $\alpha_P$, and $g(\alpha_P)$. $\alpha_P$ has been obtained from Eq. (16).

| Model | $f(\alpha)$ | $g(\alpha)$ | *G(z)* | $\alpha_P$ | $g(\alpha_P)$ |
|---|---|---|---|---|---|
| n-dimensional reaction (phase boundary reactions[1,2]), R(n) | $(1-\alpha)^{n-1/n}$, $n$=1,2 or 3 | $n\left[1-(1-\alpha)^{1/n}\right]$ | $1-\left(1-\frac{z}{n}\right)^n$ | $1-\left(\frac{n-1}{n}\right)^n$ | 1 |
| Power law,[18] P(n) | $n\alpha^{n-1/n}$, $n$=1, 2, $\frac{1}{2}$, $\frac{1}{3}$ or $\frac{1}{4}$ | $\alpha^{1/n}$ | $z^n$ | 1 | 1 |
| Kolmogorov-Johnson-Mehl-Avrami,[19-24] KJMA(n) | $n\cdot(1-\alpha)\cdot[-\ln(1-\alpha)]^{n-1/n}$ for values of $n$ see ref. [25, 26] | $[-\ln(1-\alpha)]^{1/n}$ | $1-\exp[-z^n]$ | 1-e$^{-1}$ | 1 |
| Mampel, 1st order,[27] F1 | $(1-\alpha)$ | $[-\ln(1-\alpha)]$ | $1-\exp[-z]$ | 1-e$^{-1}$ | 1 |
| 2nd order, F2 | $(1-\alpha)^2$ | $\frac{1}{1-\alpha}-1$ | $\frac{z}{1+z}$ | 0.5 | 1 |
| nth-order rate eq., F(n) | $(1-\alpha)^n$ | $\frac{(1-\alpha)^{1-n}-1}{n-1}$ | $1-[1+(n-1)z]^{1/1-n}$ | $1-n^{1/1-n}$ | 1 |
| Prout-Tompkins,[28] B1 | $\alpha(1-\alpha)$ | $\ln\left(\frac{\alpha}{1-\alpha}\right)$ | $\frac{1}{1+\exp[-z]}$ | 0.8240 | 1.5437 |
| 1D diffusion,[29] D1 | $(2\alpha)^{-1}$ | $\alpha^2$ | $\sqrt{z}$ | 1 | 1 |
| 2D diffusion,[30] D2 | $1/[-\ln(1-\alpha)]$ | $(1-\alpha)\ln(1-\alpha)+\alpha$ | - | 0.8336 | 0.5352 |
| 3D diffusion, Jander's eq.,[31] D3 | $\frac{3(1-\alpha)^{2/3}}{2\left[1-(1-\alpha)^{1/3}\right]}$ | $\left[1-(1-\alpha)^{1/3}\right]^2$ | $-3z+(3+z)\sqrt{z}$ | $\frac{19}{27}$ | $\frac{1}{9}$ |
| 3D diffusion, Ginstling and Brounshtein eq.,[32] D4 | $\frac{3}{2}\left[(1-\alpha)^{-1/3}-1\right]^{-1}$ | $1-\frac{2}{3}\alpha-(1-\alpha)^{2/3}$ | $\frac{3}{8}\left[1-4z+2\sqrt{1-24z}\sin\left(\frac{\theta+\pi/2}{3}\right)\right]$ $\theta=\arctan\left[\frac{8\sqrt{3z(1-3z)^3}}{1-60z-72z^2}\right], \theta>0$ | 0.7757 | 0.1137 |
| Šesták-Berggren,[2] SB(n,m,p) | $(1-\alpha)^n\alpha^m\left[-\ln(1-\alpha)\right]^p$ | | | | |

**Table II**. Fitted and theoretical values of the slope and y-intercept for the plot of the full width at half maximum (FWHM), $\Delta t_{FWHM}$ (in s), of the transformation rate peak versus 1000/$T_P$.

| Model | Fitted | | Theoretical | |
|---|---|---|---|---|
| | slope | y-intercept | slope | y-intercept |
| F(n), *n*=0.177 | 23340 K | -15.65 | 23360 K | -15.69 |
| SB(n,m,p), *m*=0.63, *n*=1.39, *p*=0 | 11593 K | -14.55 | 11595 K | -14.86 |
| KJMA(n), *n*=3 | 5000 K | -15.81 | 5000 K | -15.81 |
| SB(n,m,p), *n*=0.49, *m*=1.4, *p*=0 | 20537 $K^3$ | 2.50 | 20256 $K^3$ | 2.04 |

**Table III**. Fitted and theoretical values of the slope and y-intercept for the Kissinger plot (Appendix A). $\beta$ in K/s.

| Model | Fitted | | Theoretical | |
|---|---|---|---|---|
| | slope | y-intercept | slope | y-intercept |
| F(n), *n*=0.177 | -23310 K | 5.68 | -23360 K | 5.80 |
| SB(n,m,p), *m*=0.63, *n*=1.39, *p*=0 | -11556 K | 4.98 | -11595 K | 5.43 |
| KJMA(n), *n*=3 | -4984 K | 6.94 | -5000 K | 7.09 |
| SB(n,m,p), *n*=0.49, *m*=1.4, *p*=0 | 19431 $K^3$ | -12.7 | 20256 $K^3$ | -11.7 |

**Table IV**. Parameters related to the accuracy of the approximate solution, Eqs. (A.2) and (A.3) Appendix A, for the two extreme values of the heating rate (0.005 and 100 K/min).

| | $-k(T_S)/(\beta x_S')$ | | $g(\alpha_S)$ | $\frac{\Delta t'_{FWHM}}{g(\alpha_S)}$ |
|---|---|---|---|---|
| Model | $\beta = 0.005$ | $\beta = 100$ | | |
| F(n), *n*=0.177 | 1.05 | 1.08 | 1 | 1.18 |
| SB(n,m,p), *m*=0.63, *n*=1.39, *p*=0 | 1.97 | 2.03 | 1.8387 | 0.94 |
| KJMA(n), *n*=3 | 1.07 | 1.10 | 1 | 0.81 |
| SB(n,m,p), *n*=0.49, *m*=1.4, *p*=0 | 1.25 | 4.00 | 1.03 | 1.24 |

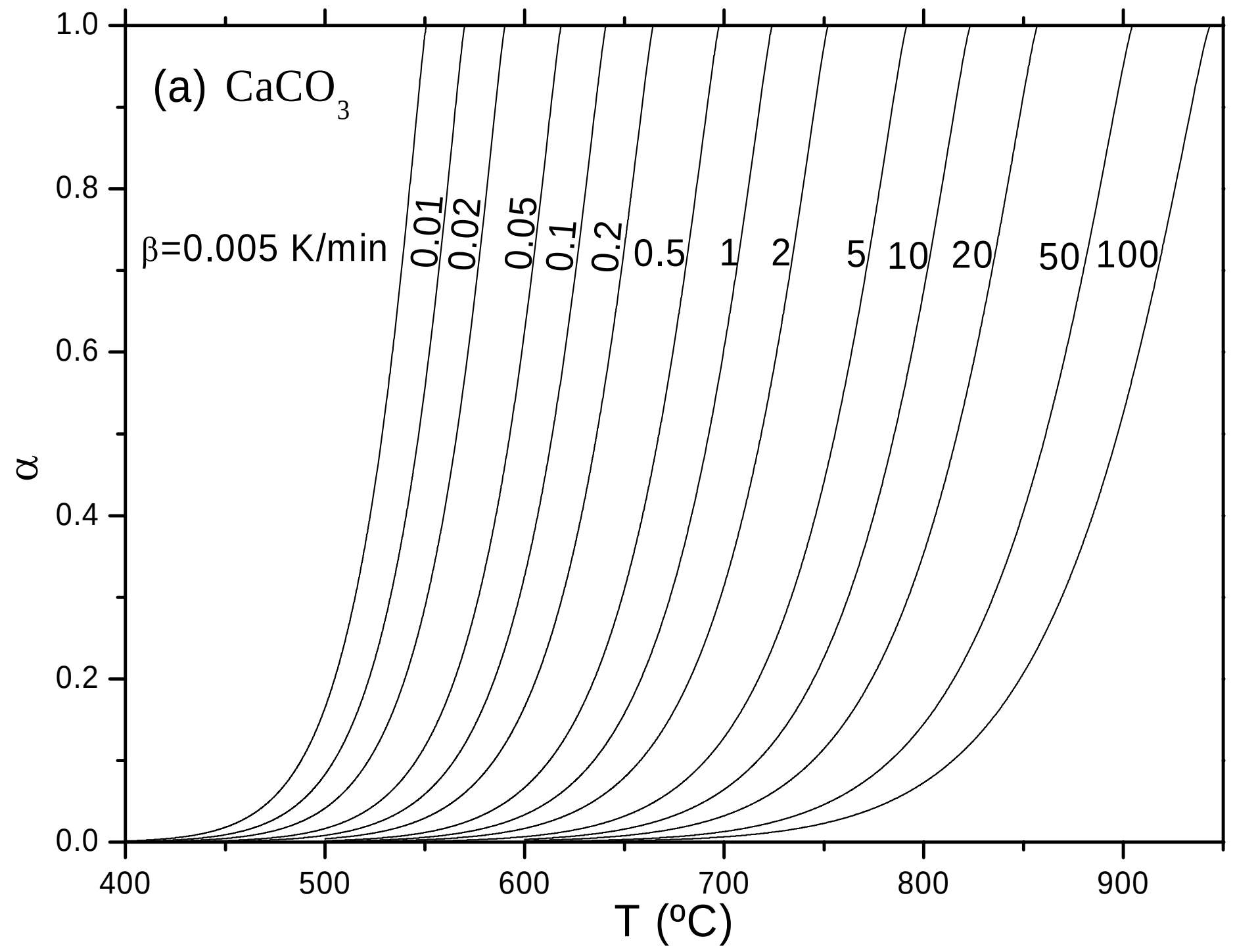

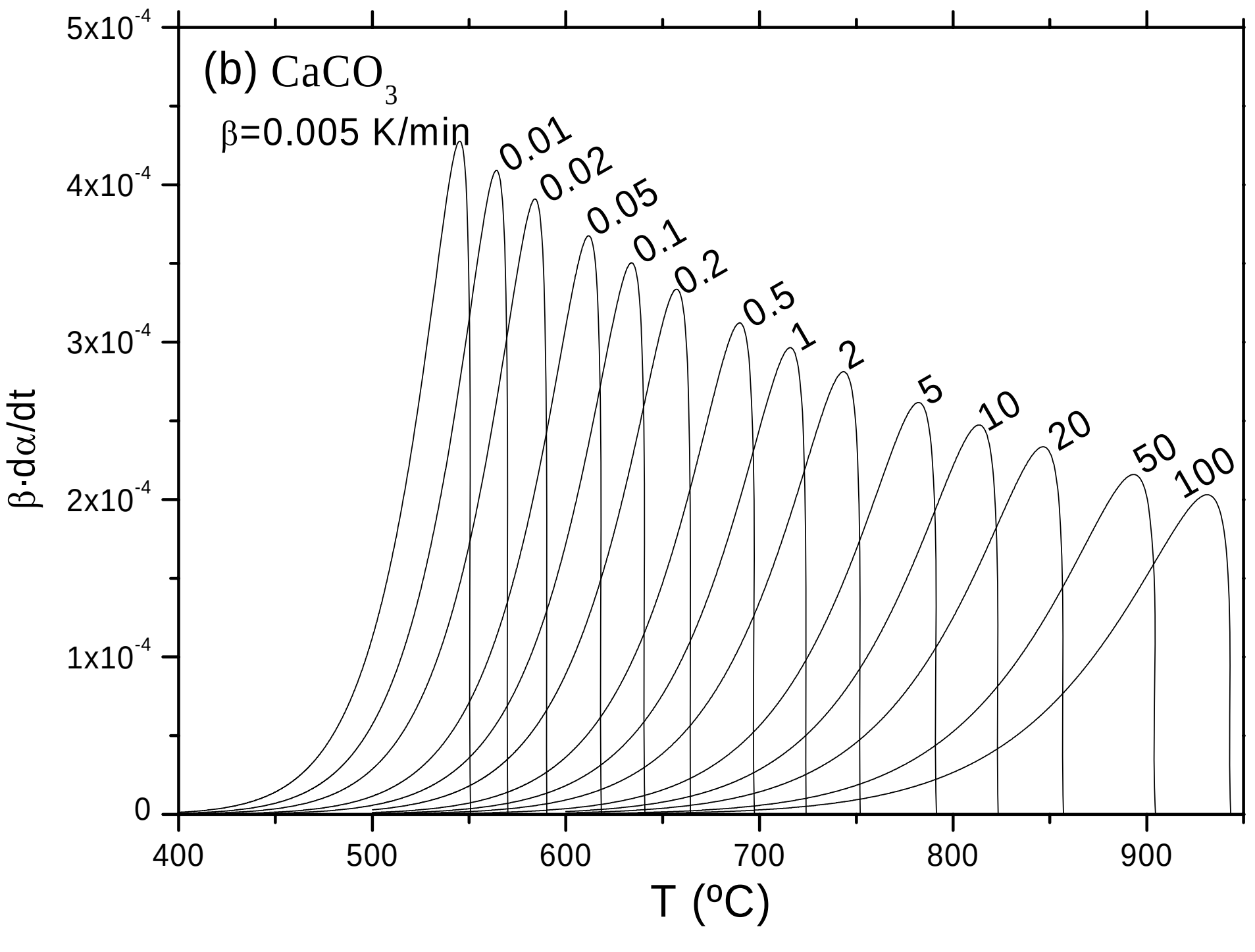

Figure 1.

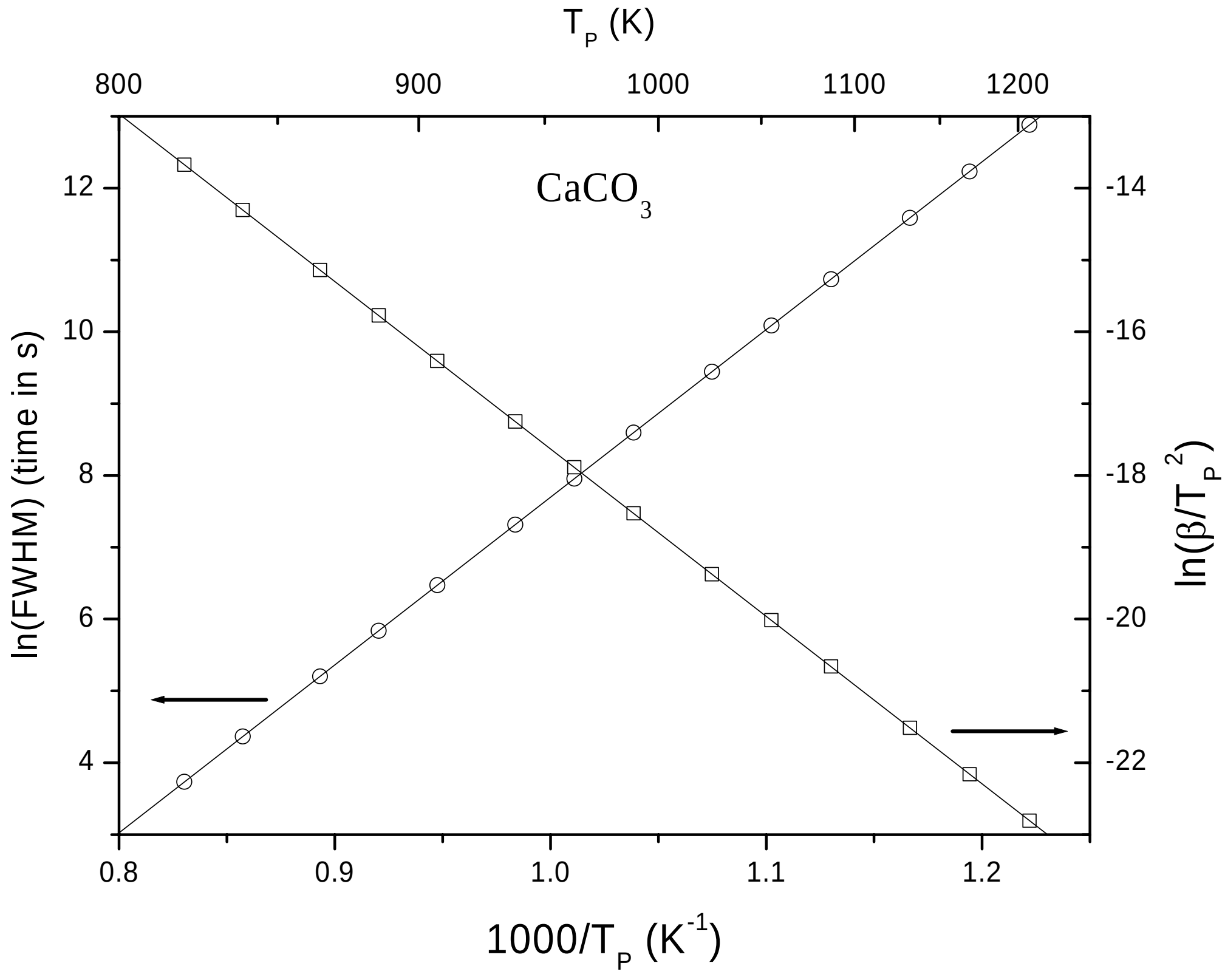

Figure 2.

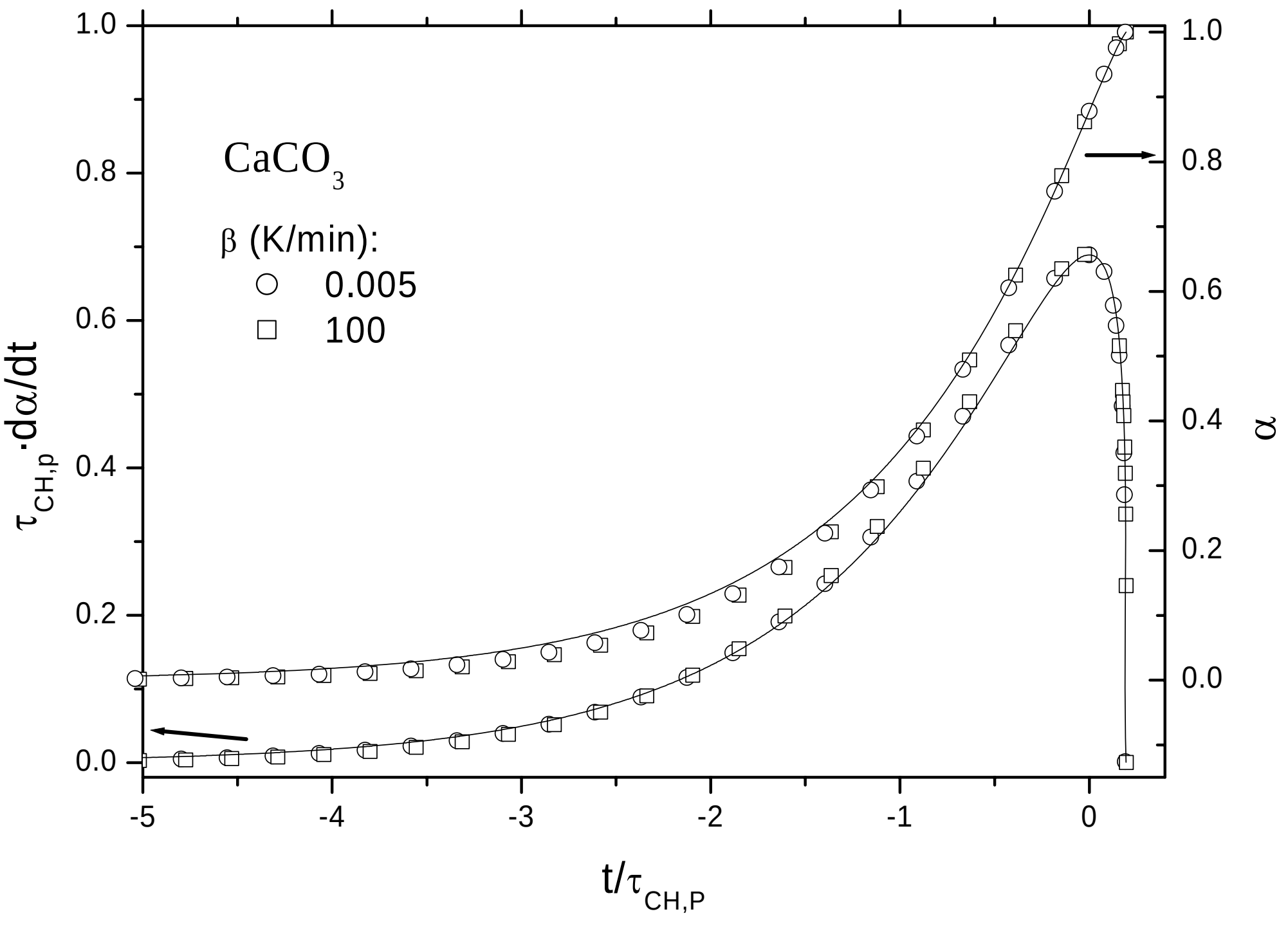

Figure 3.

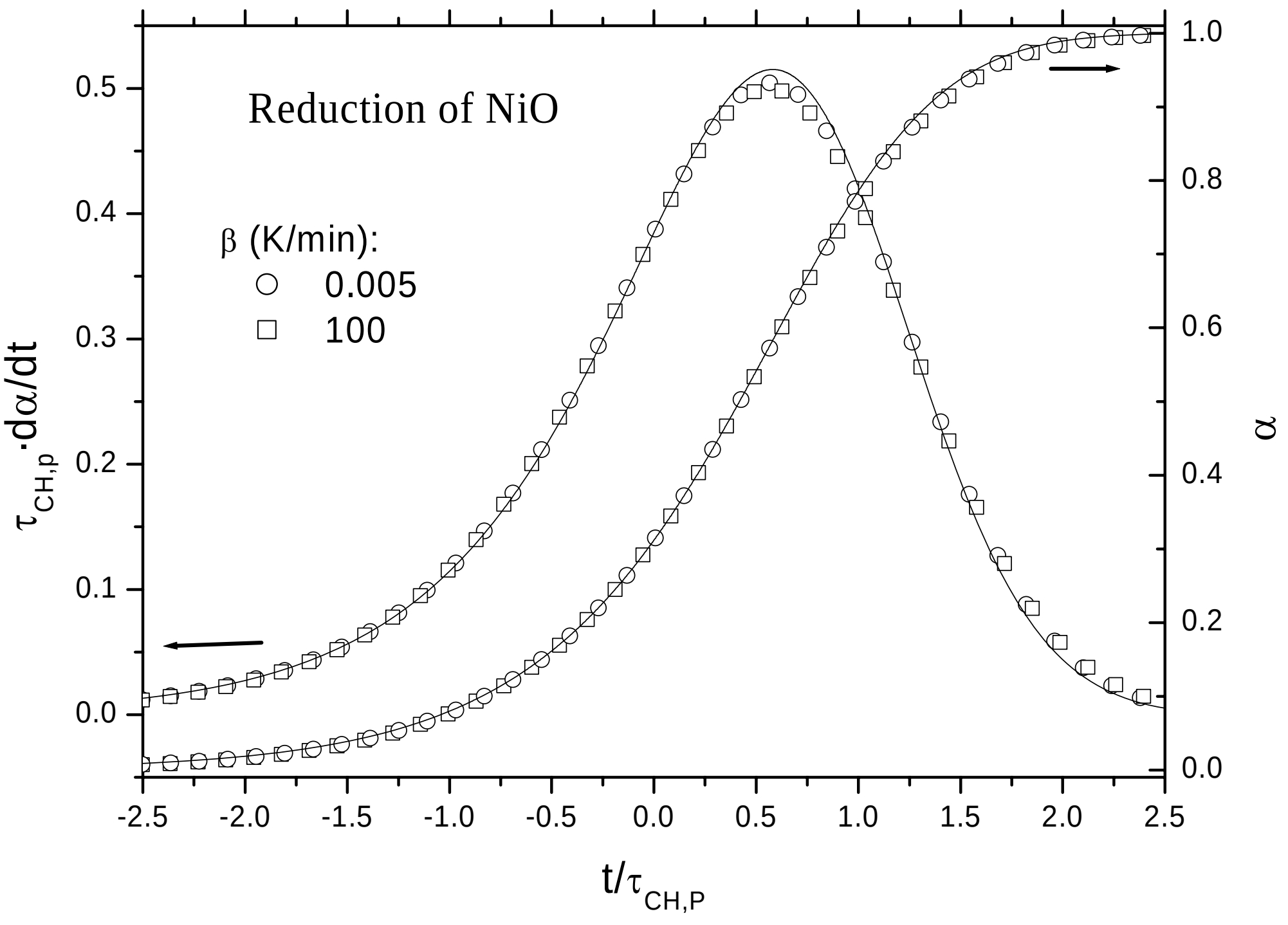

Figure 4.

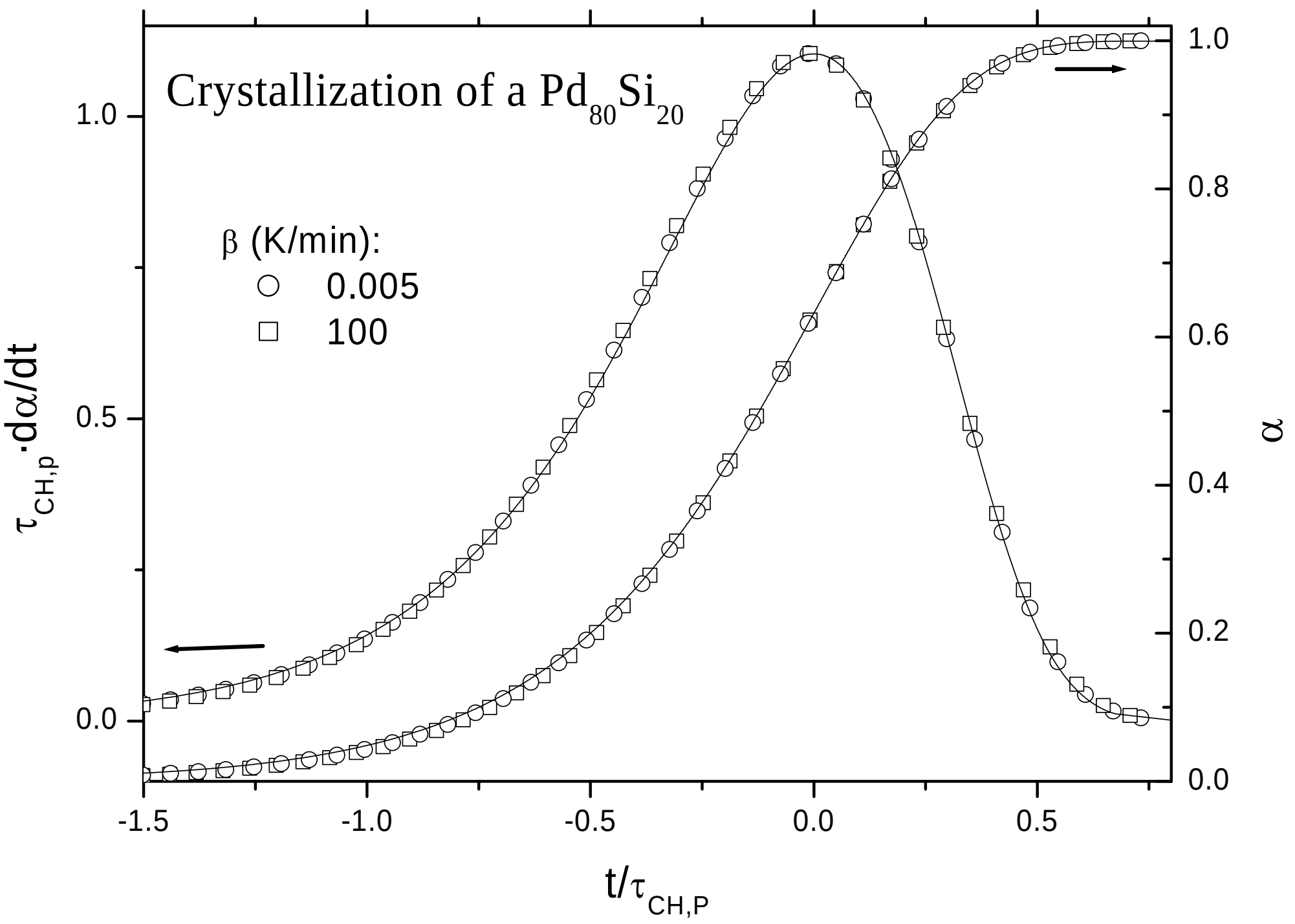

Figure 5.

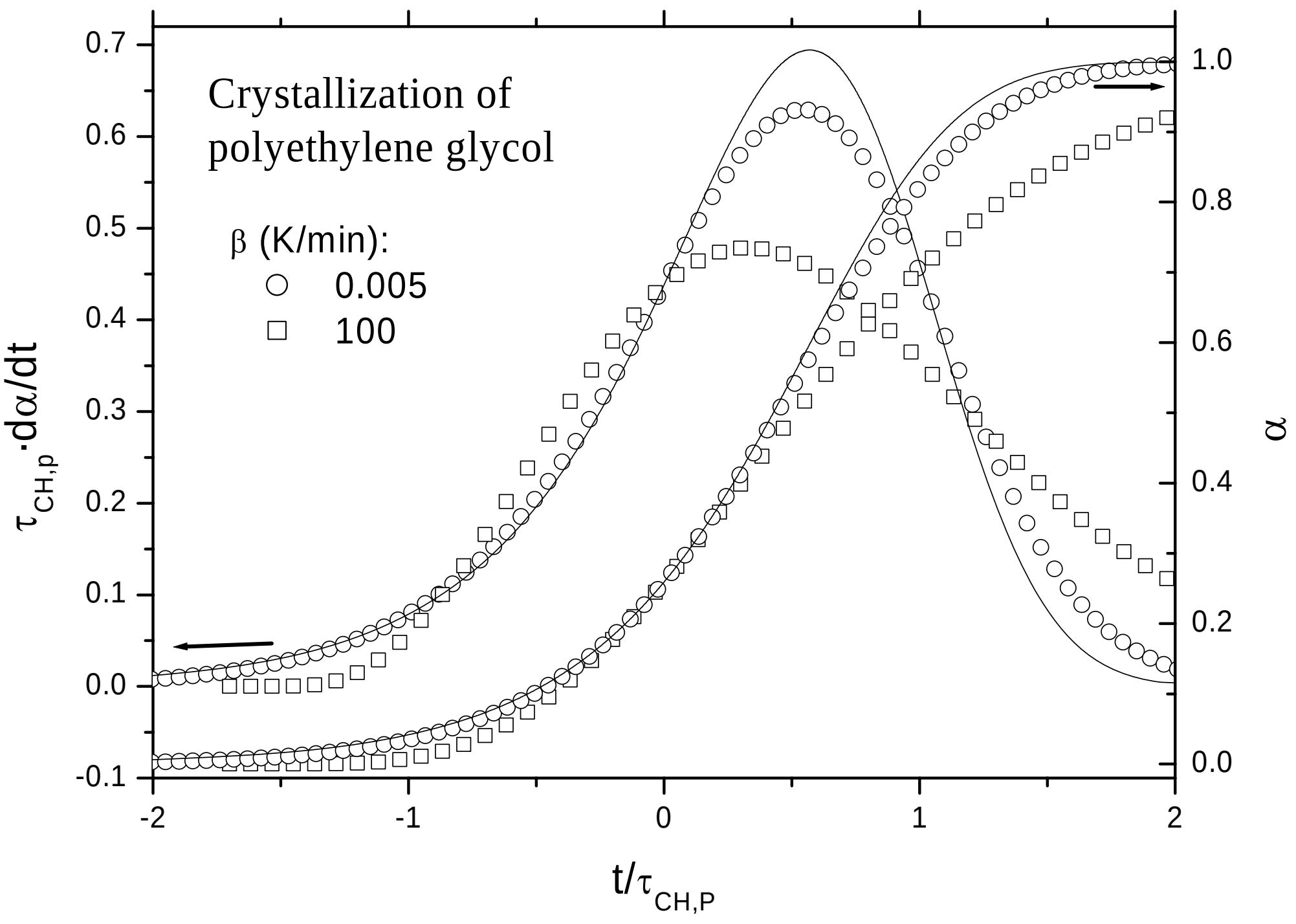

Figure 6.

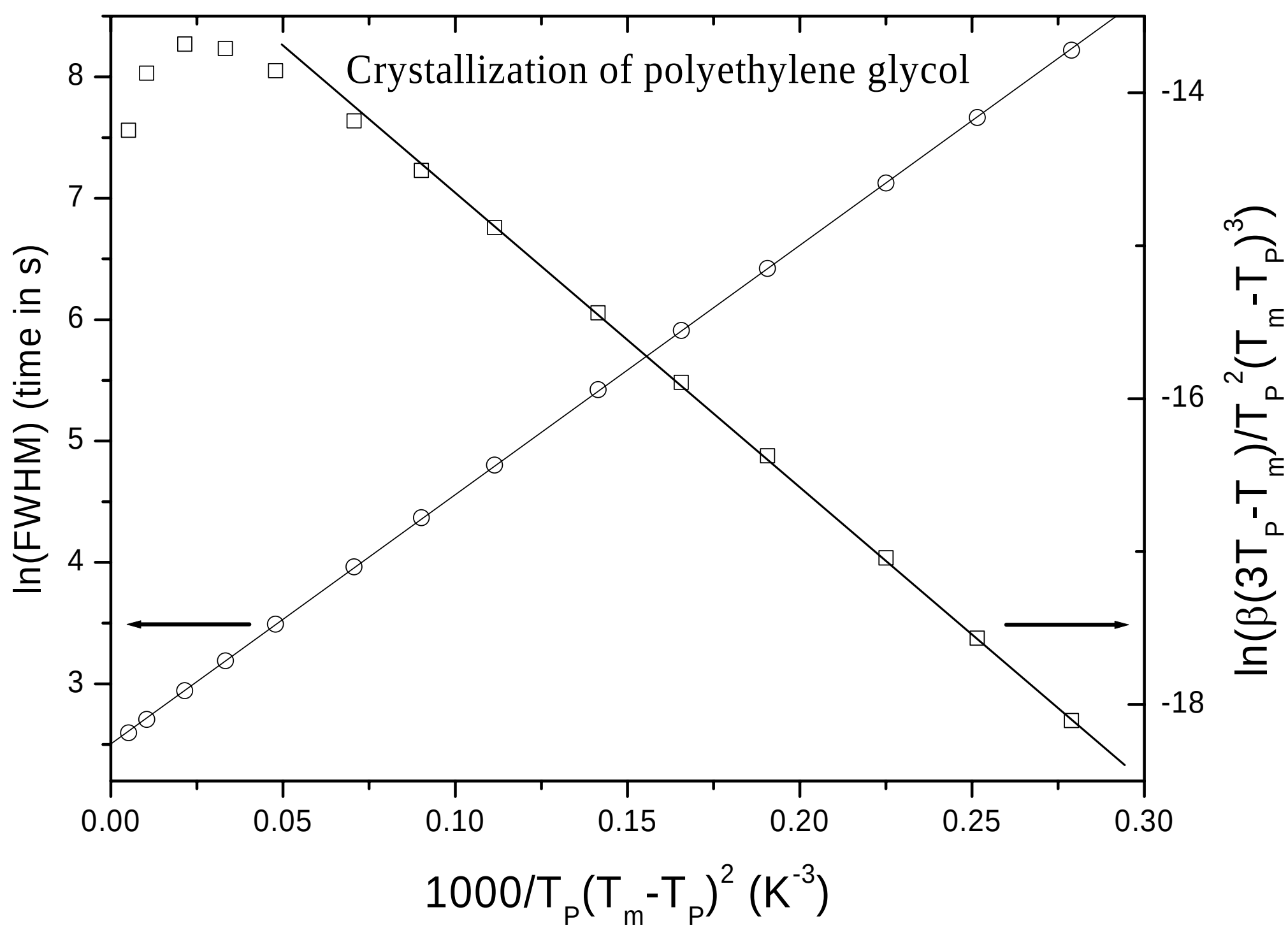

Figure 7.

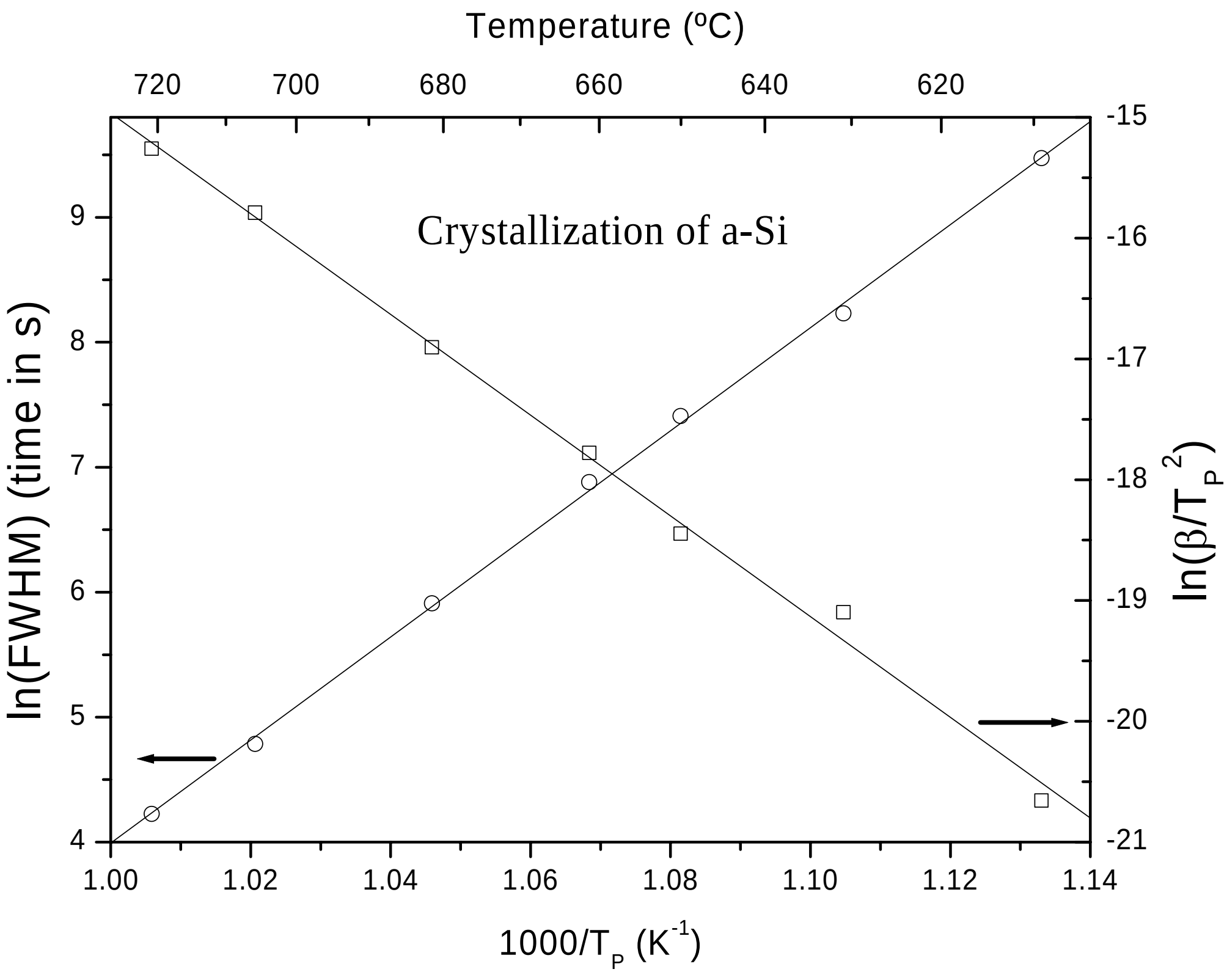

Figure 8.

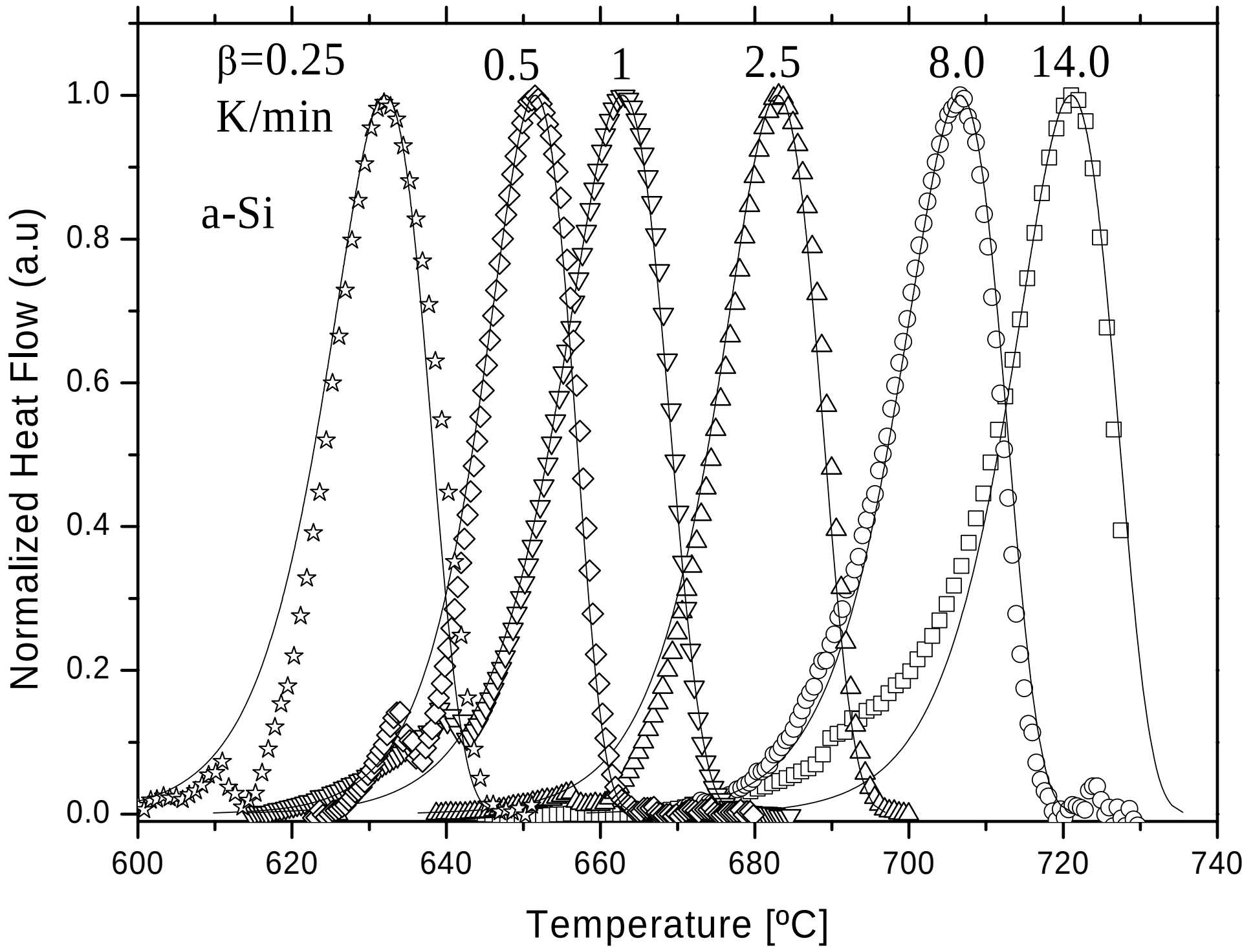

Figure 9.